\documentclass[rnote]{aa} 
\usepackage{graphicx}
\usepackage{txfonts}
\begin{document}
\title{Model-Independent Diagnostics of Highly Reddened Milky Way Star
Clusters: Age Calibration}

\author{Yuri Beletsky
       \inst{1}
       \and
       Giovanni Carraro
       \inst{1,2}
       \and
       Valentin D. Ivanov
       \inst{1}
       }

\institute{$^1$ESO, Alonso de Cordova 3107, Santiago de Chile, Chile\\
             $^2$Dipartimento di Astronomia, Universit\`a di Padova,
                 Vicolo Osservatorio 2, I-35122 Padova, Italy\\
           \email{ybialets-gcarraro-vivanov@eso.org}
            }

\date{Received June , 2009; accepted , 2009}


\abstract
{The next generation near- and mid-infrared Galactic surveys will
yield a large number of new highly obscured star clusters. Detailed characterization
of these new objects with spectroscopy is time-consuming.}
{Diagnostic tools that will be able to characterize clusters
based only on the available photometry will be needed to study
large samples of the newly found objects.}
{The brightness difference between the red clump and the main-sequence
turn-off point have been used as a model-independent age calibrator
for clusters with ages from a few 10$^8$ to 10$^{10}$\,yr in the optical. Here we
apply for the first time the method in the near-infrared.}
{We calibrated this difference in $K$-band, which is likely to be
available for obscured clusters, and we apply it to a number of test
clusters with photometry comparable to the one that will be yielded
by the current or near-future surveys.}
{The new calibration yields reliable ages over the range of ages
for which the red clump is present in clusters. The slope of the
relation is smoother than that of the corresponding $V$-band
relation, reducing the uncertainty in the age determinations with
respect to the optical ones.}


\titlerunning{Ages of Obscured Star Clusters}
\authorrunning{Beletsky et al.}

\maketitle
%

\section{Introduction}

The recent advance in the infrared (IR) instrumentation made it
possible to carry out large scale sky surveys. The Two Micron
All Sky Survey (2MASS; Skrutskie et al. \cite{skr06}) was the
first one to provide a deep look into the most obscured regions
of the Milky Way. A number of new clusters were found in 2MASS
(i.e. Dutra \& Bica \cite{dut00}, Ivanov et al. \cite{iva02b},
Bica et al. \cite{bic03}, Borissova et al. \cite{bor03}).
Currently, the UKIRT Infrared Deep Sky Survey (UKIDSS; Lawrence
et al. \cite{law07}) is creating a map of selected Northern sky
regions, typically $\sim$5\,mag deeper than the 2MASS, and in
the near future the Visible and Infrared Survey Telescope for
Astronomy (VISTA; Emerson et al. \cite{eme06}, Dalton et al.
\cite{dal06}) will provide similar coverage in the South. These
projects include specialized multicolor Galactic plane surveys:
UKIDSS Galactic plane surveys (GPS; Lucas et al. \cite{luc08})
and VISTA variables in the Via Lactea (VVV; Minniti et al.
\cite{min06}) which are expected to reveal a large number of
new star clusters.

Their characterization is a time-consuming process, requiring
spectroscopy in the near-IR, because of the strong intervening
dust absorption (A$_K$$\sim$0.1$\times$A$_V$; Rieke \& Lebofsky
\cite{rie85}). This prompted us to attempt to develop various
methods to determine the cluster parameters from the photometry
alone. For example, the metal abundances of globular clusters is
easily measured with an error of $\sigma$([Fe/H])$\sim$0.2\,dex
(i.e. Ferraro et al. \cite{fer99}, Ivanov \& Borissova
\cite{iva02}) from the slope of the RGB on the color-magnitude
diagrams (CMD).

In this {\it Note} we focus on a new age calibrator based on
the K-band. difference between the clump and the turn off point (TO)
in the CMD of star clusters older than about 300 Myr, which
will be widely used to get a quick and approximate age values
for newly discovered IR star clusters.

The layout of the paper is as follows. In Section~2 we introduce
the concept of age calibrator as derived and used in the optical
regime. Section~3 describes the sample
of very well know star clusters on which we base our new age calibrator. 
A few analytical relations are 
given in Section~4. These relations and then applied
to a few test-cases in Section~5. Section.~6, finally,
summarizes out findings.

\section{Age calibrations for open star clusters}
The magnitude difference $\Delta V$ between the main sequence (MS)
TO and the clump of He burning stars is a well
known age indicator for Galactic open clusters. The very
requirement to have a well-defined red clump limits it
to clusters older than $\sim$300\,Myr when the He burning stars
appear. However, the method is distance and reddening independent,
as demonstrated early on by Cannon (\cite{can70}), and it is not
surprising that it has been widely used ever since (see, for
instance, Carraro \& Chiosi \cite{car94}, Salaris et al.
\cite{sal04}).

Different, more sophisticated  versions of this calibrator have been
devised in the past, in the optical regime, by Anthony-Twarog \&
Twarog (\cite{ant85}) and Phelps et al. (\cite{phe94}). They also
include the color difference between the TO and the red giant
branch (RGB). However, their application is more limited because
usually the RGBs of open clusters are poorly populated, and the
RGB colors are ill-defined.

The simple difference $\Delta V$ between the red clump and TO --
the two major features in the CMD -- has the advantage of providing
a quick age estimate even from un-calibrated CMDs (Phelps et al.
\cite{phe94}, Salaris et al. \cite{sal04}). As described by Carraro
\& Chiosi (\cite{car94}) this indicator shows a slight metallicity
dependence in the optical: $\Delta V$ at fixed age is larger for
higher metallicity clusters than for lower metallicity ones.

The aim of the present work is to extend this calibrator to the near-IR,
where the metallicity effect should be weaker. We will use the
2MASS $K_S$ filter as the least affected by the extinction, and
throughout the {\it Note} will mark it as $K$, for simplicity.

\section{Sample Selection and Indicator Definition}

We selected 14 open clusters with 2MASS photometry, with reliable
age and metallicity estimates, spanning a wide range of the
parametric space (Table\,\ref{Table:sample}). The data come from
the latest edition of the
WEBDA\footnote{http://www.univie.ac.at/webda/navigation.html}
database maintained by E. Paunzen at Vienna University. Throughout
the paper we adopted a conservative estimates of the uncertainty
in the age of 15\% and a factor of 2 for the abundance [Fe/H].

\begin{table}
\caption{Parameters of the clusters' sample. The last column gives a
reference to the literature source of the adopted cluster age
and metallicity.}
\label{Table:sample}
\begin{center}
\begin{tabular}{l@{}c@{}c@{}c@{}c@{}c@{}r@{}}
\hline
\multicolumn{1}{c}{Name} &
\multicolumn{1}{c}{$RA$}  &
\multicolumn{1}{c}{$DEC$}  &
\multicolumn{1}{c}{$Age$} &
\multicolumn{1}{c}{$[Fe/H]$} &
\multicolumn{1}{c}{$\Delta K$} &
\multicolumn{1}{c}{$Ref.$}\\
\hline
& {\rm~~~$hh$:$mm$:$ss$~}
& {\rm $^{o}$~:~$^{\prime}$~:~$^{\prime\prime}$}
& [Gyr]
&
& [mag]
& \\
\hline
NGC\,188  & 00:47:28 &   +22.384 & 6.0 &   +0.01$\pm$0.08~~ & 4.0$\pm$0.15 &  1\\
NGC\,752  & 01:57:41 & $-$23.254 & 1.6 & $-$0.16$\pm$0.10~~ & 2.7$\pm$0.20 &  2\\
NGC\,2420~& 07:38:23 &   +19.634 & 2.1 & $-$0.26$\pm$0.07~~ & 3.3$\pm$0.20 &  3\\
NGC\,2477 & 07:52:10 & $-$05.850 & 0.8 &   +0.07$\pm$0.03~~ & 1.9$\pm$0.25 &  4\\
NGC\,2506 & 08:00:01 &   +09.935 & 1.9 & $-$0.37$\pm$0.02~~ & 3.1$\pm$0.25 &  5\\
NGC\,2509 & 08:00:48 &   +05.820 & 1.2 &   +0.00$\pm$0.10~~ & 2.5$\pm$0.20 &  6\\
M\,67     & 08:51:18 &   +31.896 & 4.0 &   +0.00$\pm$0.01~~ & 3.8$\pm$0.10 &  7\\
NGC\,3680 & 11:25:38 &   +16.918 & 1.8 & $-$0.16$\pm$0.03~~ & 2.9$\pm$0.15 &  8\\
NGC\,6134 & 16:27:46 & $-$00.200 & 1.0 &   +0.18$\pm$0.10~~ & 2.0$\pm$0.15 &  9\\
NGC\,6253 & 16:59:05 & $-$06.260 & 3.0 &   +0.36$\pm$0.07~~ & 3.7$\pm$0.15 & 10\\
IC\,4651  & 17:24:49 & $-$07.907 & 1.1 & $-$0.11$\pm$0.01~~ & 2.1$\pm$0.20 & 11\\
NGC\,6819 & 19:41:18 &   +08.481 & 2.5 &   +0.09$\pm$0.03~~ & 3.4$\pm$0.25 & 12\\
NGC\,6939 & 20:31:30 &   +12.304 & 2.2 &   +0.02$\pm$0.05~~ & 3.2$\pm$0.15 & 13\\
NGC\,7789 & 23:57:24 & $-$05.385 & 1.4 & $-$0.10$\pm$0.05~~ & 2.4$\pm$0.20 & 14\\
\hline
\end{tabular}\\
\end{center}
 1) Fornal et al. \cite{for07};
 2) Anthony-Twarog \& Twarog \cite{ant06b};
 3) Anthony-Twarog et al. \cite{ant06a};
 4) Kassis et al. \cite{kas97};
 5) Marconi et al. \cite{mar97};
 6) Carraro \& Costa \cite{car07};
 7) Sarajedini et al. \cite{sar09};
 8) Anthony-Twarog \& Twarog \cite{ant04};
 9) Bruntt et al. \cite{bru99};
10) Twarog et al. \cite{twa03};
11) Meibon et al. \cite{mei02};
12) Kalirai et al. \cite{kal01};
13) Andreuzzi et al. \cite{and04};
14) Vallenari et al. \cite{val00}
\end{table}

Some of the clusters suffer from significant fore- and
background contamination since they are located low
onto the Galactic plane. To minimize it, we built up
homogeneous CMDs considering only the stars within the
accepted cluster radii estimates from Dias et al.
(\cite{dia02}). The CMDs for M\,67 and NGC\,2506 -- two
clusters with rather different ages and abundances -- are shown in
Fig.~1.

\begin{figure}
\centering
\includegraphics[width=\columnwidth]{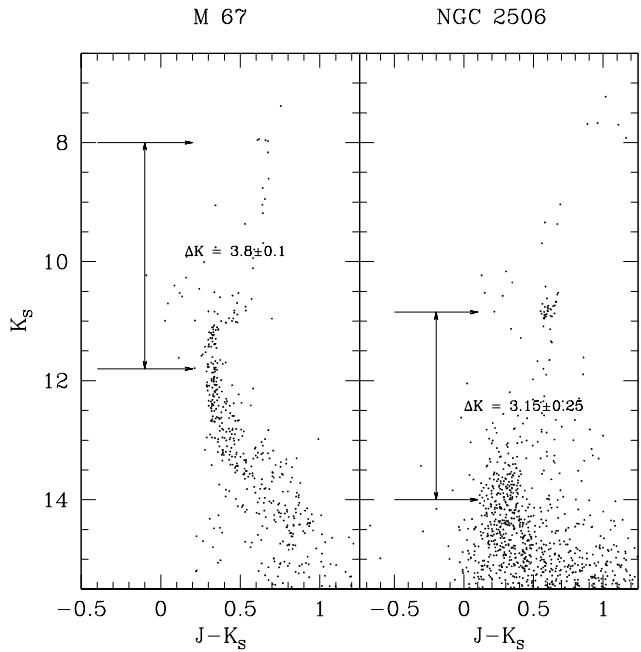}
\caption{Definition of the parameter $\Delta K$. The left
panel shows the CMD of M\,67 (solar metallicity, 4\,Gyr old),
the right shows the CMD of NGC\,2506 (significantly subsolar
metallicity, twice younger than M\,67).}
\label{Fig:CMDs}
\end{figure}

\begin{figure}
\centering
\includegraphics[width=\columnwidth]{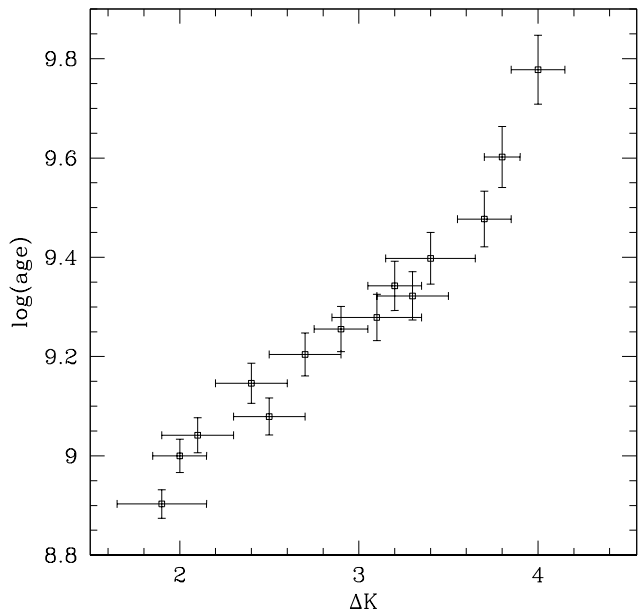}
\caption{Age vs. $\Delta K$ relation for the clusters listed in
Table\,\ref{Table:sample}.In the same Table we report the uncertainties
in age and $\Delta K$.}
\label{Fig:relation}
\end{figure}

Next, we identified the clump and TO magnitudes, respectively
at $K^{clump}$=8.00$\pm$0.05, and $K^{TO}$=11.80$\pm$0.10\,mag,
for M\,67. This implies $\Delta K$=3.8$\pm$0.1\,mag. For
NGC\,2506 we found $K^{clump}$=10.85$\pm$0.15, and
$K^{TO}$=14.00$\pm$0.20\,mag, yielding
$\Delta K$=3.15$\pm$0.25\,mag. These examples illustrate that
for an old cluster like M\,67 the definition of the clump
magnitude is easier despite the smaller number of He burning
stars, while the clump at younger ages is wider both in color
and magnitude due to evolutionary effects (Girardi et al. 2000).

The measured values of $\Delta K$ are also listed in
Table\,\ref{Table:sample}. Their uncertainties were obtained
tentatively, by eye. These take into account the broadening
of features in the CMD due to photometric errors, patchy extinction
and unresolved binaries.  
The experience shows that such estimates
can be even more reliable than the automated ones (i.e. see
the discussion in Section.~4 of Phelps \& Janes 1994). The
age (logarithm) versus $\Delta K$ relation is plotted in
Fig.\,\ref{Fig:relation}.
\\

\noindent
We emphasize that this IR relation looks smoother than the optical
counterpart (Carraro \& Chiosi 1994) mainly because
of the significantly weaker metallicity dependence (see below).
Furthermore, 
the smooth points distribution in Fig.~2 makes any
analytical fit through the data  easier and more robust .

\section{Calibration}

An analytical representation of the age versus $\Delta K$
relation is needed to facilitate the derivation of cluster
ages. The relation appears non-linear, prompting us to fit
a higher order polynomial:

\begin{equation}
log(Age)=9.09(0.29)+0.24(0.20)\times\Delta K+0.10(0.03)\times(\Delta K)^2
\label{Egn:2}
\end{equation}
with r.m.s.=0.28. The reduced $\chi^2$ of the fit is 0.08.

\begin{figure}
\centering
\includegraphics[width=\columnwidth]{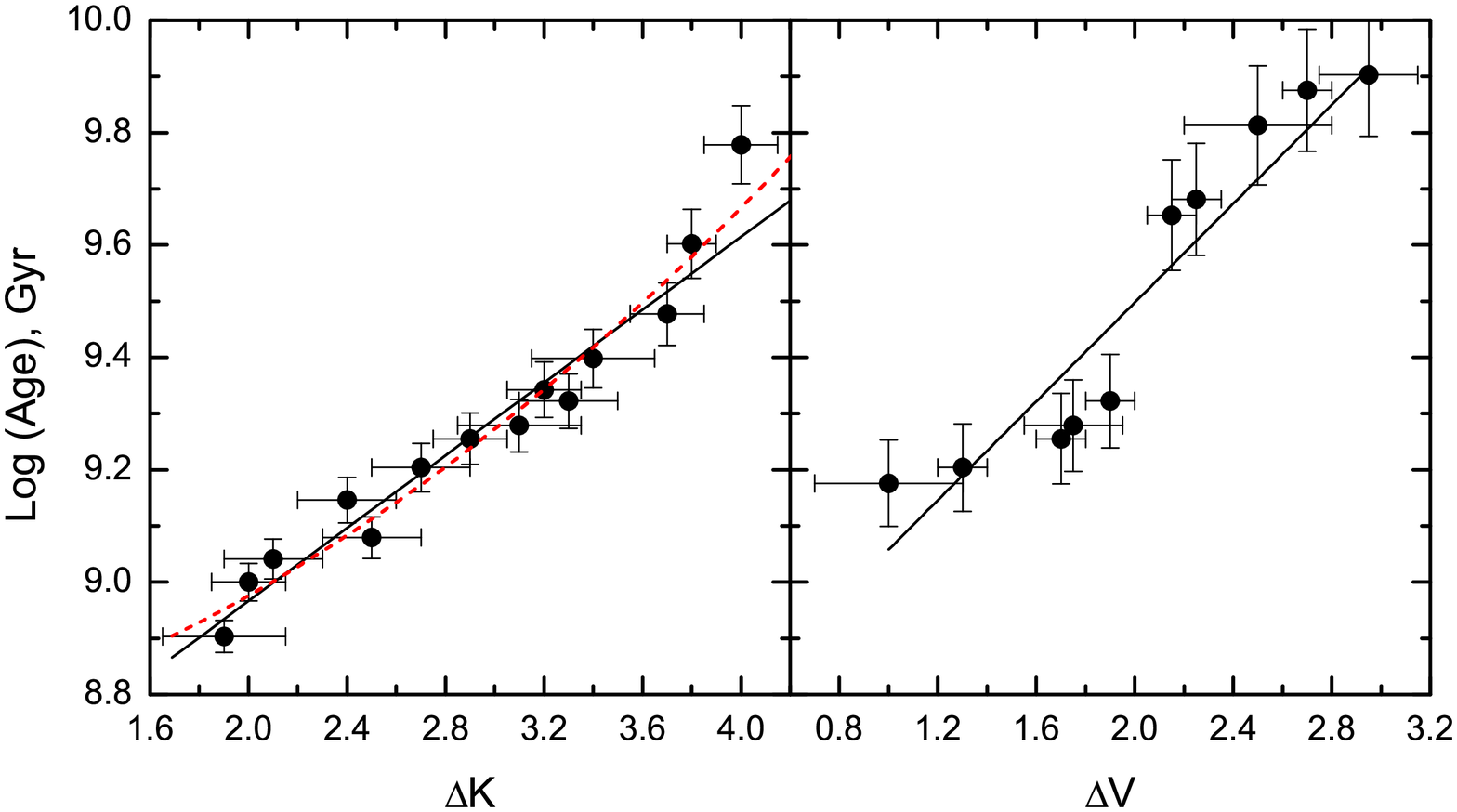}
\caption{{\bf Left panel:}Age vs $\Delta K$ relation for the 14 clusters listed
in Table\,\ref{Table:sample}.
The polynomial fit (Eqn.\,\ref{Egn:1}) is showed as a dashed
line, and the linear one (Eqn.\,\ref{Egn:2}) -- as a solid line. {\bf Right panel:} 
The optical relation from Carraro
\& Chiosi (1994). The solid line is a linear fit through the data.}
\label{Fig:fits}
\end{figure}

For comparison, the linear fit
to the data is:
\begin{equation}
log(Age)= 8.27(0.1)+0.35(0.03)\times\Delta K
\label{Egn:1}
\end{equation}
with r.m.s.=0.36. Numbers in brackets are corresponding
uncertainties.The reduced $\chi^2$ of this fit is 0.04.\\
The age in these relations is
expressed in Gyrs. The two fits are nearly indistinguishable
for 2.5$\leq$$\Delta K$$\leq$3.5\,mag -- the deviations from
the linearity occur in fact at the extremes of the age range. The
fits are plotted in the left panel of Fig.\,\ref{Fig:fits}, while in the right panel
we show the optical relation from Carraro \& Chiosi (1994), for the sake
of comparison.

As anticipated in the previous Section, the data point distribution in the IR
plot (left panel) is smoother than
the optical one (right panel), and allow a better fit through the points.

\begin{figure}
\centering
\includegraphics[width=\columnwidth]{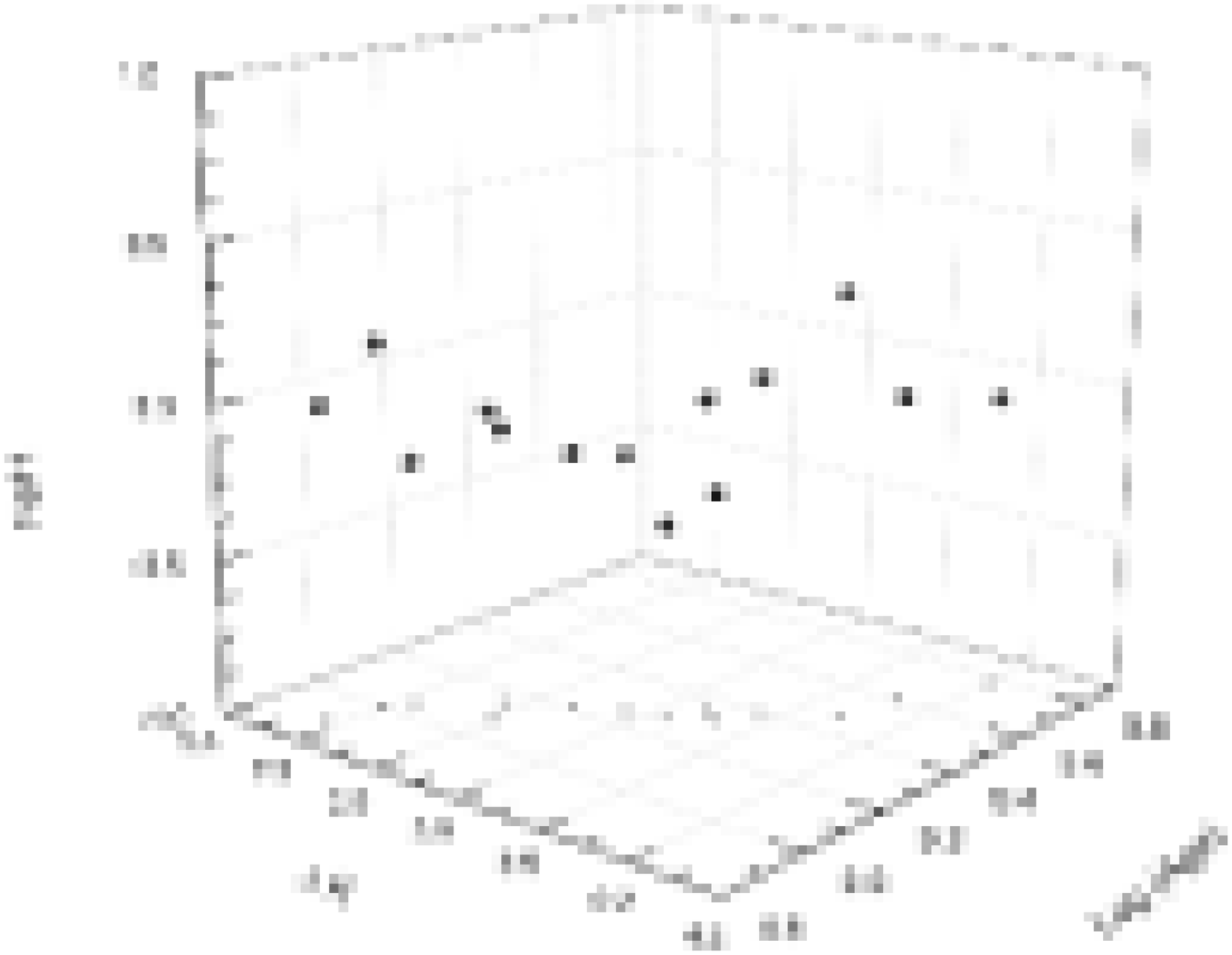}
\caption{Age vs $\Delta K$ for the 14 clusters listed in
Table\,\ref{Table:sample}, including the metallicity dependence.
The projection on the age vs $\Delta K$ is also shown.}
\label{Fig:fits_3D}
\end{figure}

Next, we include the metallicity in the fit, obtaining a
3-dimensional relation:
\begin{equation}
log(Age)=8.28(0.08)+0.34(0.03)\times\Delta K+0.00005(0.1)\times[Fe/H]
\label{Egn:3}
\end{equation}
with r.m.s.=0.06 (Fig.\,\ref{Fig:fits_3D}).
The reduced $\chi^2$ of the fit is 0.004.

As expected, the dependence on metallicity is negligible, making
$\Delta K$ very useful, since in most cases the newly discovered
clusters will not have metallicity estimates available. In
comparison with the optical relation (Carraro \& Chiosi
\cite{car94}, Eqn.~2), the linear relation in the IR is less sensitive
to metallicity, and also shows a somewhat
shallower dependence on $\Delta K$ (0.34 versus 0.45, see Carraro
\& Chiosi 1994).\\

\noindent
As for the uncertainty on the age, it is easy to show - by simple
propagation - that in
the linear cases (Eqns. 2 and 3), the uncertainty in the age
does not depend on  the value of $\Delta K$. This is not the case of course
for Eqn.~2, which has a quadratic dependence on  $\Delta K$.

\section{Validation of the new calibrations}

To demonstrate the power of the new calibrations, we applied them
to estimate the ages of three clusters
with shallow 2MASS photometry, but that have IR
photometry from other sources: NGC\,6791, NGC\,2141, and
Berkeley\,17. Spectroscopic metallicity estimates are also
available for them, allowing us to use all three relations
in a  metallicity range which encompasses the typical metal
content of Galactic open clusters. 
The results are listed in Table\,\ref{table:tests}. The age
uncertainties were derived propagating the measurement errors
through the calibrations.

\subsection{NGC\,6791}

IR photometry for this cluster is taken from Carney et al.
(\cite{car05}). It was acquired with the IRIM camera at
the 4-m Mayall telescope at Kitt Peak National Observatory.
The data were photometrically calibrated with the UKIRT
standards (Casali \& Hawarden \cite{cas92}), which were also
used to calibrate the 2MASS itself. Not surprisingly, Eqn.\,37
from Carpenter (\cite{carp01}):
\begin{equation}
K_{2MASS}=K_{UKIRT}+0.004(0.006)\times(J-K)_{UKIRT}+0.002(0.004)
\end{equation}
indicates that the color difference $\Delta (J-K)$$\sim$0.3\,mag
between the TO and clump leads to only 0.0012$\pm$0.0018\,mag
difference between the K-band magnitudes of the two systems --
much smaller than our measurement errors. The constant term is
negligible for calculating the $\Delta K$ parameters, and it is
statistically indistinguishable from zero, anyway.

\begin{figure*}
\centering
\includegraphics[width=18cm]{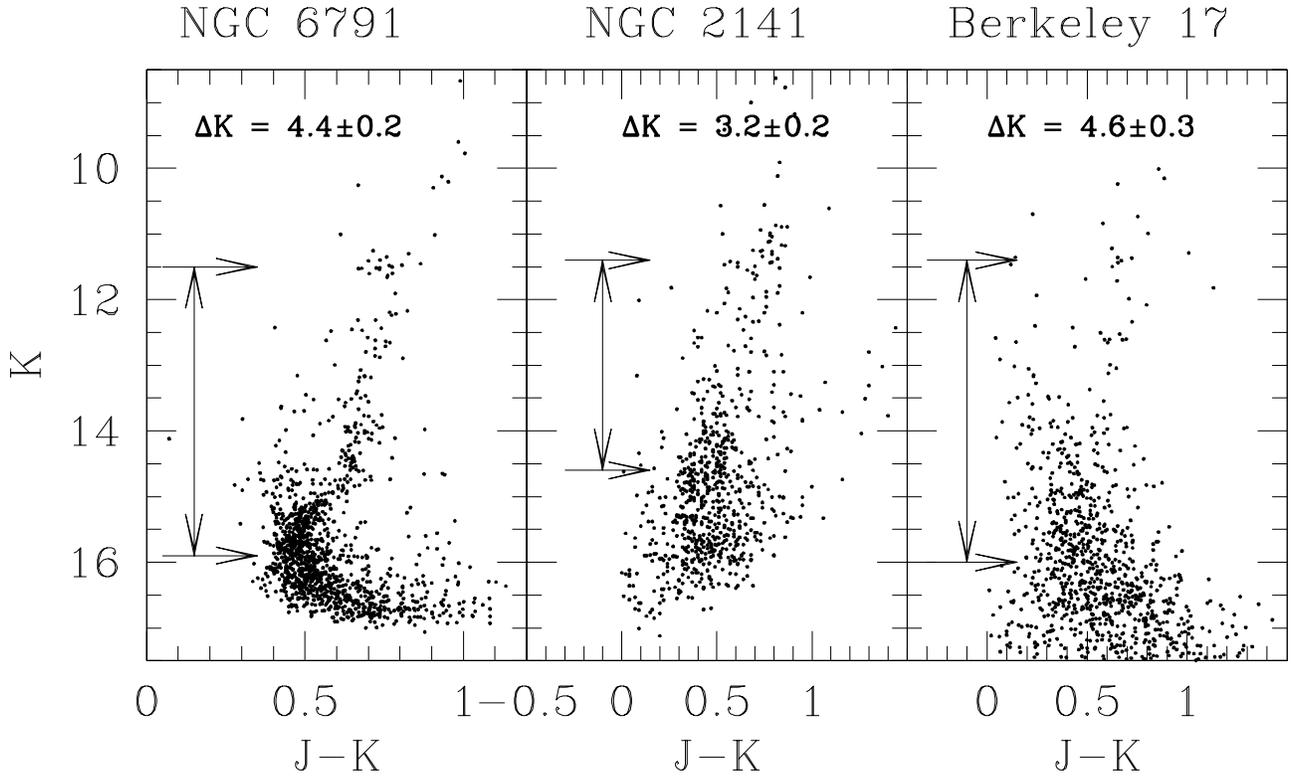}
\caption{Application of the new age indicator to NGC\,6791,
NGC\,2141 and Berkeley~17 (from left to right). $\Delta K$ is
marked on each CMD.}
\label{Fig:aplications}
\end{figure*}

The CMD of NGC\,6791 is shown in the left panel of
Fig.\,\ref{Fig:aplications}. Widely accepted values for age and
metallicity are in the range of 7-9 Gyrs and [Fe/H]=+0.35 to
+0.45 dex, respectively(Carney et al. 2005,  and references therein). 
Our equations 1,
2, and 3, yield ages of 9.3, 6.5 and 6.0\,Gyrs, respectively.

\subsection{NGC\,2141}

IR photometry for this cluster is taken from Carraro et al.
(\cite{car01}). It was acquired with the ARNICA camera at the
Tirgo telescope. The data were photometrically calibrated with
the ARNICA standards (Hunt et al. \cite{hun98}). Eqn.\,8 from
Carpenter (\cite{carp01}):
\begin{equation}
K_{2MASS}=K_{ARNICA}-0.024(0.011)\times(J-K)_{ARNICA}+0.012(0.006)
\end{equation}
indicates that the color difference $\Delta (J-K)$$\sim$0.25\,mag
between the TO and clump leads to only $\sim$ 0.006$\pm$0.003\,mag
difference between the K-band magnitudes of the two systems --
again, much smaller than our measurement errors.

The CMD of NGC\,2141 is shown in the middle panel of
Fig.\,\ref{Fig:aplications}. For this cluster, metallicity
is [Fe/H]=$-$0.26, and the age is $\sim$2.5\,Gyrs  (Carraro et al. 2001). 
We obtained ages of 2.2, 2.4 and 2.3\,Gyrs,
respectively from Eqns. 1, 2, and 3. These results are in
good agreement with the isochrone fitting estimates
(Carraro et al. \cite{car01}).

\subsection{Berkeley\,17}

IR photometry for this cluster is taken from Carraro et al.
(\cite{car99}). It was acquired with the ARNICA camera at the
Tirgo telescope, and photometrically calibrated with the
ARNICA standards (Hunt et al. \cite{hun98}). As shown in the
previous subsection, our age measuring parameter $\Delta K$
is not affected significantly by using a different photometric
system than the 2MASS.

The CMD of Berkeley\,17 is shown in the right panel of
Fig.\,\ref{Fig:aplications}. This cluster is known to have
marginally subsolar abundance [Fe/H]=$-$0.10 dex and is it
9-10\,Gyrs old (Friel et al. 2005). Our calibrations yielded
ages of 10.0, 7.6 and 7.0\,Gyrs, respectively for Eqns. 1, 2,
and 3.

\vspace{3mm}
This exercise shows that the age of a cluster within the linear
regime of our calibrator (2.5$\leq$$\Delta K$$\leq$3.5\,mag) is
closer to the isochrone-derived one when simple linear (Eqn.\,1),
or the metal dependent (Eqn.\, 3) fitting are used. The most
deviating relation is the polynomial one. In the other two cases
of very old clusters, the polynomial relation (Eqn.\, 2) is
the most accurate, and provides age estimates very close to the
widely accepted in the literature.
Finally, the new calibrations provide,   
in the linear regime, age estimates that agree quite well with
isochrone-based ages, while outside this range the difference might
be of the order of 30\%.

\begin{table*}
\caption{New age estimates for the three test-cases discussed in
this paper. The last three columns list the ages determined from
Eqns.\, 1, 2 and 3, respectively. References for the
literature ages and abundances are given in the text.}
\label{table:tests}
\begin{center}
\begin{tabular}{@{}l@{}c@{}c@{}c@{}c@{}c@{}c@{}c@{}c@{}}
\hline
\multicolumn{1}{c}{Name} &
\multicolumn{1}{c}{$RA$}  &
\multicolumn{1}{c}{$DEC$}  &
\multicolumn{1}{c}{$Age$} &
\multicolumn{1}{c}{$[Fe/H]$} &
\multicolumn{1}{c}{$\Delta K$} &
\multicolumn{1}{c}{$Age (1)$} &
\multicolumn{1}{c}{$Age (2)$} &
\multicolumn{1}{c}{$Age (3)$} \\
\hline
& {\rm ~$hh$:$mm$:$ss$~}
& {\rm $^{o}$~:~$^{\prime}$~:~$^{\prime\prime}$}
& [Gyr]
&
& [mag]
& [Gyr]
& [Gyr]
& [Gyr] \\
\hline
NGC\,6791    & 19:20:53 & $+$37:46:18  & 8.0 & $+$0.40 & 4.40$\pm$0.20 &  9.3$\pm$1.3 & 6.5$\pm$0.5 & 6.0$\pm$0.5 \\
NGC\,2141    & 06:02:55 & $+$10:26:48  & 2.5 & $-$0.26 & 3.20$\pm$0.20 &  2.2$\pm$0.2 & 2.4$\pm$0.2 & 2.3$\pm$0.2 \\
Berkeley\,17~& 05:20:36 &~$+$30:36:00~ & 9.0 & $-$0.10 & 4.60$\pm$0.20 & 10.0$\pm$1.4 & 7.6$\pm$0.6 & 7.0$\pm$0.7 \\
\hline
\end{tabular}
\end{center}
\end{table*}

\section{Summary and conclusions}

In this paper we described relations of the magnitude difference
between the TO and the clump $\Delta K$ in the IR CMD versus age for
800-1000\,Myr old open star clusters. The age limits are set by
the life time of the red clump stars and available photometric data.
Our goal is to develop a simple tool to measure the age of newly
detected star clusters from the next generation IR sky surveys.
The relations are derived in the widely used 2MASS photometric
system. We provide linear and second order polynomial fits of the
age versus $\Delta K$ relationship, and a multi-linear fit which
also takes into account the metallicity of the clusters, when available.
However, it is unlikely that the metallicity will be immediately
available for most of the new objects, so Eqns.\,1 and 2 will be mostly used 
used for {\it en-masse} analysis of the new clusters. 

Beside, we show that the dependence on metallicity is very shallow,
which makes these relations of more robust use than their
optical counterpart.

We then illustrate
with three examples that the simple linear relation works better
for intermediate $\Delta K$ values, while the polynomial relation
is preferable at the extremes of the validity range.\\

\noindent
The method can potentially be useful for a wider age range, and
these calibrations should be extended in the future to cover both
younger and older clusters.

\begin{acknowledgements}
This research has made use of the SIMBAD database, operated
at CDS, Strasbourg, France, and of the WEBDA
database maintained by E. Paunzen at Vienna University.
\end{acknowledgements}

\end{document}